\documentclass[aps,twocolumn,showpacs]{revtex4}
\usepackage{epsfig}

\begin{document}

\title{Event synchronization: a simple and fast method to measure synchronicity and 
     time delay patterns.}

\author{R. Quian Quiroga$^{1, 2}$}
   \altaffiliation{Corresponding author. E-mail: rodri@vis.caltech.edu}
\author{T. Kreuz$^{1, 3}$ and P. Grassberger$^1$}
\affiliation{$^1$ John von Neumann Institute for Computing, Forschungszentrum J\"ulich,
   D-52425 J\"ulich, Germany\\
$^2$ Sloan-Swartz Center for Theoretical Neurobiology, Div. of Biology, 139-74,\\
   California Institute of Technology, Pasadena, 91125, CA\\
$^3$ Department of Epileptology, University of Bonn, Sigmund-Freud Str. 25,
   D-53105 Bonn, Germany}

\date{\today}

\begin{abstract}

We propose a simple method to measure synchronization and time delay patterns between 
signals. It is based on the relative timings of events in the time series, defined 
e.g. as local maxima. The degree of synchronization is obtained from the number 
of quasi-simultaneous appearances of events, and the delay is calculated from the 
precedence of events in one signal with respect to the other.
Moreover, we can easily visualize the time evolution of the delay and synchronization 
level with an excellent resolution. 

We apply the algorithm to short rat EEG signals, some of them containing spikes. We also 
apply it to an intracranial human EEG recording containing an epileptic seizure, and 
we propose that the method might be useful for the detection of foci and for seizure 
prediction. It can be easily extended to other types of data and it is very simple 
and fast, thus being suitable for on-line implementations.

\end{abstract}

\maketitle

\section{Introduction}

In recent years, several measures of synchronization have been proposed and applied 
successfully to different types of data. 
Among these studies we can distinguish two main approaches: 
1) One based on similarities of trajectories in phase space (constructed e.g. by 
time-delay embedding) \cite{schiff,quyen,arnhold,quian,quian1}; 
2) One that measures phase differences between the signals, where the phases are 
defined either from a Hilbert \cite{rosemblum,tass,florian} or from a wavelet 
transform \cite{lachaux,rodriguez} (as shown in \cite{quian1}, these 
two apparently different phases are indeed closely related).

These new methods compete in popularity with standard measures such as the 
cross-correlation, the coherence function, mutual information, and also with simple
visual inspection of the recordings. 
Cross-correlation and coherence are clearly the measures most used so far. 
In contrast to them, all new measures are {\it non-linear} in the sense that they 
depend also on properties beyond second moments. In addition, some of them have
the advantage of being asymmetric, eventually being able to show driver-response 
relationships \cite{arnhold,quian}.

Among others, synchronization measures have been used for the study of 
electroencephalogram (EEG) signals. Applications include prediction and localization 
of epileptic activity \cite{quyen,arnhold,florian}, phase-locking between different 
recording sites upon visual stimulation \cite{lachaux,rodriguez}, resonance between EEG and 
muscle activity in Parkinson patients \cite{tass}, desynchronization upon lesions in the 
thalamic reticular nucleus in rats \cite{sleepwake}, synchronization in motoneurons 
within the spinal cord \cite{schiff}, etc.   

In the present paper we present a very simple algorithm that can be used for any 
time series in which we can define {\it events}. These can be spikes in single-neuron
recordings, epileptiform spikes in EEGs, heart beats, stock market crashes, etc. In 
principle, when dealing with signals of different character, the events
could be defined differently in each time series, since their common cause might
manifest itself differently in each series. 
%For instance, they could be maxima in one MEG channel, but minima in another.
This {\it event synchronization} (ES) does not require the notion of phase. It cannot
distinguish between different forms 
of $m:n$ lockings \cite{rosemblum,tass}, but it can tell which of the two time series
leads the other. And, above all, it is very simple conceptually and easy to implement.
Due to that, it can be used on-line and can show rapid changes of synchronization
patterns.

\section{Event synchronization and delay asymmetry}

Given two simultaneously measured discrete univariate time series $x_n$ and $y_n$, 
$n=1,\ldots,N$, we first define suitable {\it events} and event times $t^x_i$ and 
$t^y_j$ $(i = 1, \ldots, m_x; 
j = 1, \ldots, m_y)$. In the signals to be analyzed in this paper, these events will be
simply local maxima, subject to some further conditions. If the signals are synchronized, 
many events will appear more or less simultaneously. Essentially, we count the fraction
of event pairs matching in time, and we count how often each time series leads in these
matches. Similar concepts were used in \cite{pijn}.

Let us first assume that there is a well defined characteristic event rate in each 
time series. Counter examples include strong chirps and onsets of epileptic seizures 
where event rates change rapidly. Such cases will be treated below.
Allowing a time lag $\pm\tau$ between two `synchronous' events (which should be 
smaller than half the minimum inter-event distance, to avoid double counting), 
let us denote by $c^{\tau} (x|y)$ the number of times an event appears in 
$x$ shortly after it appears in $y$, i.e:
\begin{equation}
    c^{\tau} (x|y) = \sum_{i=1}^{m_x} \sum_{j=1}^{m_y} J_{ij}^\tau         \label{eq:cxy}
\end{equation}
with
\begin{equation}
   J_{ij}^\tau = \cases{ 1     & if ~~$0 < t_i^x - t_j^y \leq \tau$ \cr
                           1/2   & if ~~$ t^x_i = t^y_j $ \cr
                        0        & else}                                   \label{eq:j}
\end{equation}
and analogously for $c^{\tau} (y|x)$. Next, we define the symmetrical and anti-symmetrical
combinations\begin{equation}
   Q_{\tau} = \frac {c^{\tau}(y|x) + c^{\tau}(x|y)} {\sqrt{m_x m_y}}  \ , \quad
   q_{\tau} = \frac {c^{\tau}(y|x) - c^{\tau}(x|y)} {\sqrt{m_x m_y}}  \ ,  \label{eq:qxy}
\end{equation}
which measure the synchronization of the events and their delay behavior, respectively.
They are normalized to $0 \leq Q_{\tau} \leq 1$ and $-1 \leq q_{\tau} \leq 1$. We 
have $Q_{\tau} =1$ if and only if the events of the signals are fully synchronized. In 
addition, if the events in $x$ always precede those in $y$, then $q_{\tau} = 1$. 

In cases where we want to avoid a global time scale $\tau$ since event rates change 
during the recording, we use a local definition $\tau_{ij}$ for each event pair $(ij)$. 
More precisely, we define 
\begin{equation}
   \tau_{ij} = \min \{ t^x_{i+1} - t^x_i, t^x_i - t^x_{i-1}, 
                   t^y_{j+1} - t^y_j, t^y_j - t^y_{j-1}\} \,/2\;.          \label{eq:tauij}
\end{equation}
We then define $J_{ij}$ as in Eq.(\ref{eq:j}) with $\tau$ replaced by $\tau_{ij}$, 
and $c(x|y)$ as in Eq.(\ref{eq:cxy}) with $J_{ij}^\tau$ replaced by $J_{ij}$. The 
factor $1/2$ in the definition of $\tau_{ij}$ avoids double counting if, e.g., two
events in $x$ are close to the same event in $y$. Of course, one could also make
other choices, e.g. by taking $\tau_{ij}$ smaller than in Eq.(\ref{eq:tauij}) or by
using $\tau'_{ij} = \min\{\tau, \tau_{ij}\}$. As in the definition of events, an
optimal choice of $\tau$ depends on the problem. In the 
following we shall suppress the dependence on $\tau$, understanding that all formulas 
apply for both variants.

To obtain time resolved variants of $Q$ and $q$ we simply modify eq.(\ref{eq:cxy}) to
\begin{equation}
     c_n(x|y) = \sum_i \sum_j J_{ij}\, \Theta(n-t_i^x)
                       \label{eq:cxyt}
\end{equation}
with $n = 1, \ldots, N$ and $\Theta$ the step function (i.e. $\Theta (x) = 0$ for 
$x \leq 0$ and $\Theta(x) = 1$ for $x > 0$). Similarly, $c_n(y|x)$ is obtained by 
exchanging $x$ and $y$. Then, we define the time resolved anti-symmetric combination 
$q(n) = c_n(y|x) - c_n(x|y)$ which can be seen as a random 
walk that takes one step up every time an event in $x$ precedes one in $y$ 
and one step down if vice versa. If an event occurs simultaneously in both signals 
or if it appears only in one of them, the random walker does not move. 
Exchanging $x$ and $y$ just reverses the walk.
For non-synchronized signals, we expect to obtain a random walk with the typical 
diffusion behavior. With delayed synchronization we will have a bias going up 
(down) if $x$ precedes (follows) $y$. 
We should remark that such a bias clearly shows the presence of a time delay of one 
signal with respect to the other, but does not necessarily prove a driver-response 
relationship, although it might suggest it. In fact, internal delay loops of one of 
the systems can fool the interpretation. Also, the two signals might be driven by a 
common hidden source and the bias just indicates different delays.

The time course of the strength of ES can be obtained from $Q(n) = c_n(y|x) + c_n(x|y)$. 
If an event is found both in $x$ and $y$ within the window $\tau$ (resp. $\tau_{ij}$), 
$Q(n)$ increases one step, otherwise it does not change. Of course, $Q(n)$ will 
also not change if there are no new events at all. The synchronization level at time 
$n$, averaged over the last $\Delta n$ time steps, is thus obtained as
\begin{equation}
   Q'(n) = {Q(n)-Q(n-\Delta n)\over \sqrt{\Delta n_x\Delta n_y}}\;,
\end{equation}
where $\Delta n_x$ and $\Delta n_y$ are the numbers of events in the interval
$[n-\Delta n,n]$.
Similarly, we can also define instantaneous delay asymmetries $q'(n)$.

\section{Applications}

Let us now apply these concepts to two sets of intracranial EEG recordings, one from 
rats and the other from an epileptic patient. 

\subsection{Rat EEGs}

\begin{figure}
\begin{center}
\epsfig{file=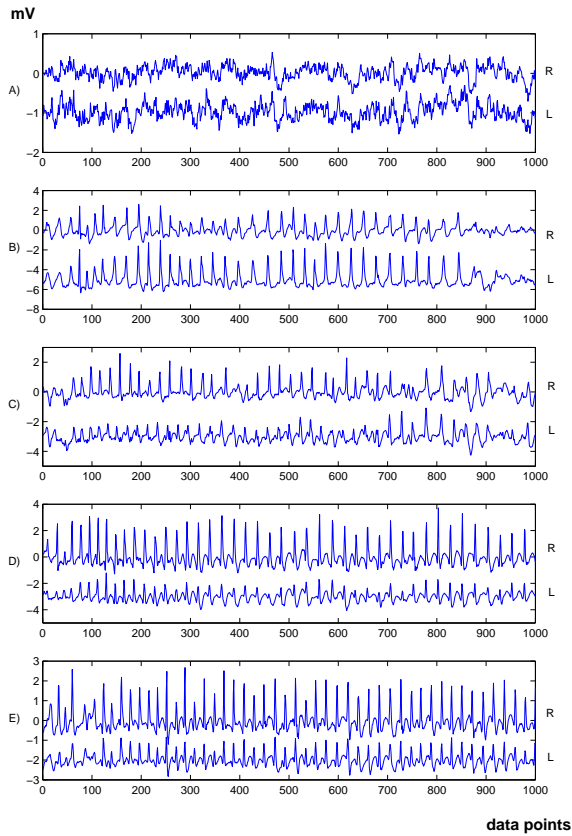, width=7.6cm,angle=0}
\end{center}
\caption{Five pairs of rat EEG signals from right and left cortical intracranial 
   electrodes. For a better visualization, left signals are plotted with an offset.}
\label{fig:examples}
\end{figure}

The five pairs of rat EEG signals were
obtained from electrodes placed on the left and right frontal cortex of male
adult WAG/Rij rats (a genetic animal model of human epilepsy) \cite{giles}. They
were referenced to an electrode placed in the cerebellum, filtered between 1-100
Hz and digitized at 200 Hz. In Fig.~\ref{fig:examples} we show these signals \cite{www}.
The first pair (example A in Fig.~\ref{fig:examples}) is a normal EEG, all others 
contain spike discharges (not to be confused with spikes in single neuron recordings)
which are the landmark of epileptic activity. They arise from abnormal synchronization 
in an epileptic brain even when there are no seizures. A localized appearance of spikes 
can indeed delimit a zone with abnormal activity (though this will not necessarily 
be the epileptic focus). Furthermore, time delays between them can identify the 
electrode closest to the epileptic focus, especially at the onset of seizures. 

Several measures of synchronization were recently applied to the first three cases 
of Fig.~\ref{fig:examples} \cite{quian1}. Since spike trains lasted usually about 
5 seconds, the challenge was to try the different measures in these short 
epochs. Surprisingly, nearly all the measures gave qualitatively similar results 
hard to be guessed beforehand. These examples and two additional cases (D and E), 
also containing spikes, will be further analyzed in this paper. 

For the example A it is difficult, due to its random-like appearance, to 
visually estimate its level of synchronization and any delay of one electrode with 
respect to the other. However, we can already observe some patterns appearing 
nearly simultaneously in both the left and right channels, thus showing some 
degree of interdependence.
The spike-wave trains in the other examples in principle suggests a high level of 
synchronization. However, as already shown in \cite{quian1}, the spikes of example 
C appear with a varying time lag between right and left channels and are therefore 
much less synchronized than those in B.
This is of course not easily seen by visual inspection of Fig.~\ref{fig:examples},
but will be clear from the following analysis.

Events were defined as local maxima fulfilling the following additional conditions: 
\begin{enumerate}
\item $x(t_i) > x(t_{i+k})$,  for $k = -K+1,\ldots, 0, \ldots,K-1$
\item $x(t_i) > x(t_{i \pm K}) + h $
\end{enumerate}
and the same for $y$. We took $K = 3$ and $h = 0.1$. Other choices gave very 
similar results.

Since the rate of events is more or less constant, we used a fixed $\tau$. The
choice $\tau=2$ gave a good discrimination between the five cases.
All results shown below were compared to those obtained with surrogate pairs
which were defined by shifting the left channel signals 500 data points (2.5 sec) 
to the right, with periodic boundary conditions. 
Our test hypothesis is that without changing the individual properties 
of each signal, after a large enough shifting synchronization should reach a 
background `zero' level. The usefulness of such surrogates was discussed in 
in more detail in \cite{quian1}. 

\begin{table}
\begin{center}
\begin{tabular}[c]{c c c c c c}
{\bf Example} & \vline & {\bf $Q_{\tau = 2}$} & {\bf $q_{\tau = 2}$} & 
           {\bf $Q_{\tau = 2}^{^{\rm surr}}$} & {\bf $q_{\tau = 2}^{^{\rm surr}}$} \\
\hline 
{\bf A} & \vline & 0.57 & ~0.15 & 0.24 & -0.01  \\
{\bf B} & \vline & 0.80 & -0.29 & 0.29 & ~0.01  \\
{\bf C} & \vline & 0.48 & -0.20 & 0.13 & -0.01  \\
{\bf D} & \vline & 0.93 & -0.59 & 0.41 & ~0.04  \\
{\bf E} & \vline & 0.90 & -0.13 & 0.46 & ~0.03  \\
\end{tabular}
\vspace{0.5cm}
\caption{Time averaged event synchronization $Q$ and delay $q$ for the five examples 
   of Fig.~\ref{fig:examples}. Positive values of $q$ indicate that events in the left 
   side lags behind the right one. Surrogate values of synchronization were obtained by 
   shifting left channel signals by 500 data points.} 
\label{tab:q}
\end{center}
\end{table} 

For the five EEG signals of Fig.~\ref{fig:examples}, we show the values of $Q_{\tau =2}$ 
and $q_{\tau =2}$ in Table~\ref{tab:q}, both for the original signals and the 
`time-shifted' surrogates. We observe that synchronization levels  rank 
${\rm D > E > B > A > C}$.
This is in agreement with the analysis of examples A, B and C done in 
\cite{quian1} with several other measures of synchronization. Note that even example A 
is ranked consistently with the other measures, although it does not contain obvious 
events such as the spikes of the other examples.

\begin{figure}
\begin{center}
\epsfig{file=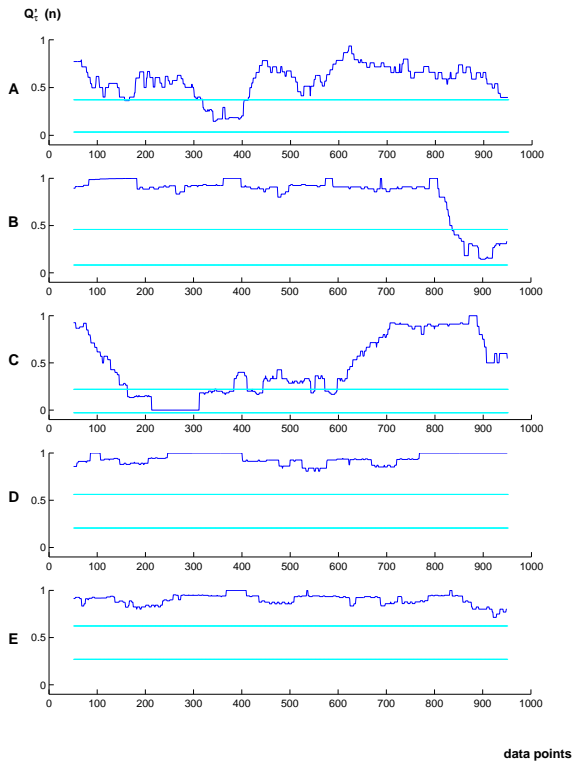, width=7.6cm,angle=0}
\end{center}
\caption{Time resolved event synchronization $Q'_{\tau = 2} (n)$ for the examples 
   of Fig.~\ref{fig:examples}. Blue horizontal lines correspond to the time averages
   $\pm 1\sigma$ of the surrogate.}
\label{fig:qs}
\end{figure}

All synchronization values are clearly higher than those of their corresponding 
surrogates (surrogates constructed with other delay values gave similar results). These 
surrogate values vary a lot for the different examples. This stresses the importance 
of keeping the individual properties of the signals when constructing surrogates. 
Except for example A, the values of $q$ show that the signals from the right hemisphere
lag behind the left ones ($y$). A closer visual inspection of Fig.~1 at higher 
resolution shows that this lag is usually 1 data point. The reason of this systematic 
lag is unclear (it could be an artifact of the data acquisition or a real physiological 
effect) and it is beyond the scope of this paper.

In Fig.~\ref{fig:qs} we show the time evolution of synchronization $Q'(n)$ for the 
five examples, calculated with a window of $\Delta n = 100$ data points. For most
of the time, they are higher than the values calculated from time-shifted surrogates 
(the light blue horizontal lines indicate time averages $\pm 1$ standard deviation).
In examples A, B and C we see abrupt changes of synchronization with time which seem 
statistically significant.
In retrospect they can also be seen in Fig.~\ref{fig:examples} on closer inspection, 
but they are much less obvious there and could easily be missed. Compared to the
first three, examples D and E are more stable in time. Finally, the time resolved 
ES shows a better resolution than all synchronization measures considered in 
\cite{quian1}.

\begin{figure}
\begin{center}
\epsfig{file=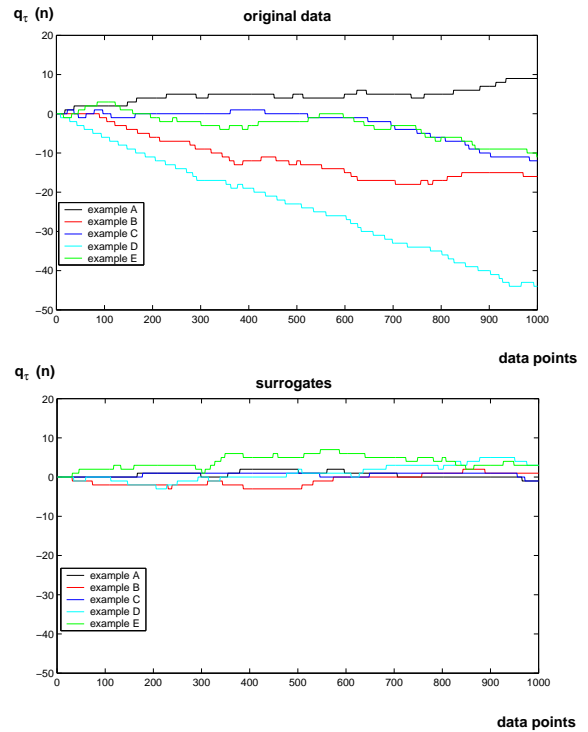, width=7.6cm,angle=0}
\end{center}
\caption{Time delays between the right and left channels (upper plot) and for the surrogates
  (lower plot). Up (down) shifts mean precedence of the right (left) channel.}
\label{fig:qa}
\end{figure}

Figure \ref{fig:qa} shows the time resolved asymmetry between the right and the 
left channels (upper plot) and the results from surrogates (lower plot). In all
five cases, the bias is in agreement with the $q$ values shown in Table~\ref{tab:q}.
The bias in example D is not only the strongest but also the most constant, confirming
that D shows the most robust and stationary ES (compare Fig.~\ref{fig:qs}).
For the other examples we see regular changes with time. This 
is of course very difficult to see in the original recordings, and it was also
not seen with any of the synchronization measures studied in \cite{quian1}.
As expected, for the surrogates we obtain random walks with small and 
erratic displacements.

\subsection{Human EEG}

\begin{figure}
\begin{center}
\epsfig{file=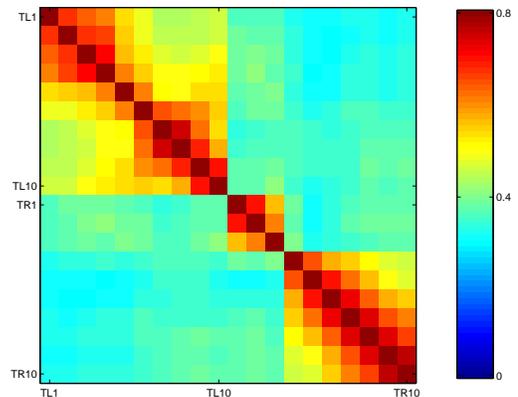, width=6.7cm,angle=0}
\end{center}
\caption{Time averaged event synchronization between the contacts on the left and right depth 
   electrodes (TL1-10 and TR1-10, respectively).}
\label{fig:matrix}
\end{figure}

As a second example we analyzed an intracranial EEG recording from an epileptic patient 
containing 12 min. of pre-seizure and seizure EEG. Data were recorded from 2 needle shaped 
depth electrodes with 10 contacts each. They were symmetrically placed in the left (contacts 
TL1 to TL10) and right (contacts TR1 to TR10) temporal lobes, in the entorhinal cortex 
and hippocampal formation. The EEG was sampled at 173~Hz and band pass filtered 
between 0.53-40~Hz. For further details on the data we refer to \cite{arnhold}. As in the 
previous example, event times were defined as local maxima, but using $K=10$ and $h=50$ 
(this large $K$ was needed because the data are more noisy than the rat data, and 
smaller values would have led to many spurious events).
Due to the varying event rate, we used a variable-$\tau$ approach. For the time 
resolved event synchronization $Q'(n)$ we took a window $\Delta n = 1730$. 

\begin{figure}
\begin{center}
\epsfig{file=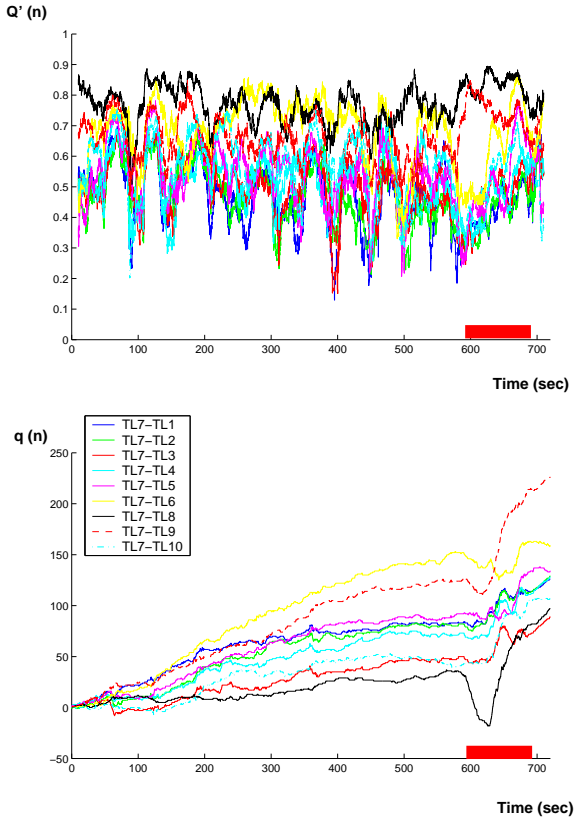, width=7.6cm,angle=0}
\end{center}
\caption{Time resolved event synchronization (upper plot) and delay asymmetries (lower plot)
   between an
   channel near the epileptic focus (TL7) and the remaining channels on the same side.
   The red bar shows the duration of the epileptic seizure.}
\label{fig:tl7}
\end{figure}

Figure \ref{fig:matrix} shows the time-averaged event synchronization values between all 
channels. A detailed analysis of synchronization patterns for similar recordings has 
already been described by Arnhold et al. \cite{arnhold} using a robust measure of 
non-linear synchronization. Here, we just summarize the main results which are in 
perfect agreement with those in Ref.~\cite{arnhold}. We first note that 
synchronization between left and right electrodes is relatively low and that the right 
contacts form two clusters: TR1-3 and TR4-10. This is just due to the fact that the 
first 3 contacts were located in the entorhinal cortex and the remaining ones 
in the hippocampus \cite{arnhold}. Moreover, for the right side we observe 
a gradual decrease of synchronization with increasing distance between contacts.
The synchronization pattern for the left channels is different. There, the entorhinal 
cortex/hippocampus separation is overshadowed by the epileptic activity leading to a 
higher overall synchronization level.

A visual analysis of the seizure onset revealed that contacts TL7 and TL8 showed the first 
signs of seizure activity.
Figure \ref{fig:tl7} shows the time resolved synchronization $Q'(n)$ and delays $q(n)$ 
between TL7 and the remaining left side channels.
As expected, synchronization is largest between TL7 and its neighbors TL8 and TL6. It 
is not homogeneous in time and we have several short drops before seizure starts. 
Moreover, starting at seizure onset and during the whole seizure, synchronization of TL7
with TL8 and TL9 is high, while synchronization with TL6 and all others is 
decreased. The lower panel shows that all left channels lag behind channel TL7. 
There is just one exception: During the first part of the seizure,
channel TL7 falls back and channel TL8 leads for about half a minute (indeed, the
lead of TL7 is weakened already some 3 minutes before the seizure). After this, 
TL7 takes up its lead even more vigorously than before. This might indicate
that the source of epileptic activity moves.
Whether these features are common to many epileptic seizures and whether they can have 
clinical significance for e.g. seizure anticipation or focus localization requires 
further study with a larger database. 

\begin{figure}
\begin{center}
\epsfig{file=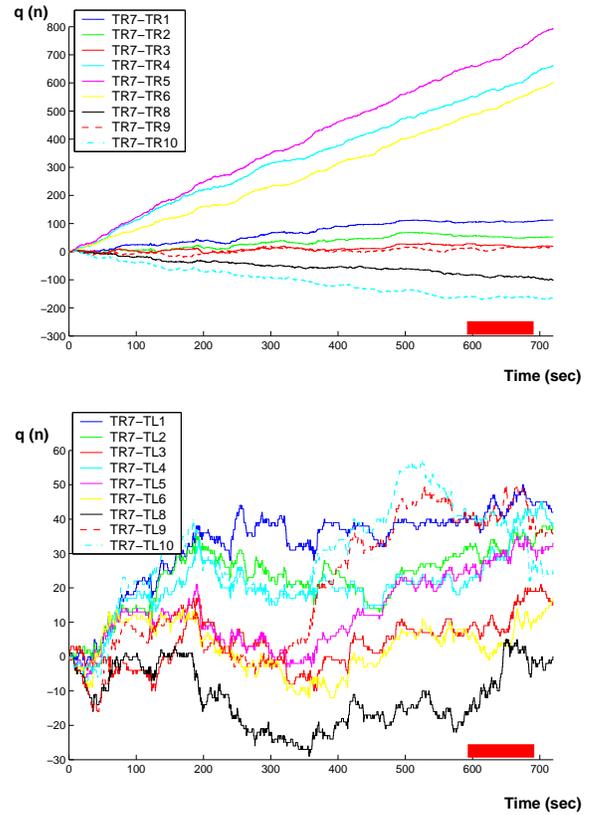, width=7.6cm,angle=0}
\end{center}
\caption{Delay patterns between an contact in the non-focal side (TR7) against the other 
contacts in the non-focal side (upper plot) and against the contacts in the focal 
side (lower plot). No anomalous behavior is seen during the seizure (red bar). Notice 
the different scales in the two plots.}
\label{fig:tr7}
\end{figure}

In Fig.~\ref{fig:tr7} we show the delays of the contralateral channel (TR7) with respect 
to the other right channels (upper plot) and to the left channels (lower plot). Channels 
TR4-6 strongly and steadily follow channel TR7, which itself follows channels TR8 and TR10. 
This might reflect the source of `normal' synchronized activity.
A detailed analysis is outside the scope of this paper and will be further addressed 
elsewhere. As seen from the lower panel, synchronization between both hemispheres is weak 
and $q$ shows unbiased random walks. The complete absence of any deviant behavior during 
the seizure reflects the fact that the seizure does not spread to the contralateral side.

\section{Conclusion}

In conclusion we presented a new approach to measure synchronization and time delays that is 
based on the relative timings of events (in this study defined as local maxima). This also 
gives an easy visualization of time-resolved synchronization and delay patterns. The method 
is appealing due to its simplicity, straightforward implementation and speed. These features 
make very easy its on-line implementation. In the particular case of EEGs, the proposed 
approach is promising for the study of recordings of epileptic patients, where 
synchronization is important and the analysis of time delay patterns could be useful for the 
localization of the epileptic focus and the prediction of seizure onset. 
Also, the method should be well suited for single-neuron recordings, where the fast
dynamics of spikes makes difficult the analysis with other measures.  
In this paper we focussed on application to EEG signals, but the method can be easily 
applied to other types of data just by adjusting the definition of events.

%\newpage
%\section*{Acknowledgments}

We are very thankful to Ralph Andrzejak, Alexander Kraskov, Klaus Lehnertz, and Heinz 
Schuster for stimulating discussions, to 
Giles van Luijtelaar and Joyce Welting from NICI, University of Nijmegen, for the rats data 
used in this paper and to K. Lehnertz and C. Elger from the Department of Epileptology, 
University of Bonn, for the intracranial EEG data. T.K. is supported by the 
Deutsche Forschungsgemeinschaft, SFB TR3.

%\newpage

\end{document}